\def\be{\begin{equation}}
\def\ee{\end{equation}}
\def\ba{\begin{array}}
\def\p{\prime}
\def\ea{\end{array}}

\def\Rb{{I\!\! R}}

\def\Cb{\ \hbox{\vrule width 0.6pt height 6pt depth 0pt
              \hskip -3.5 pt} C}
\def\cb{\ \hbox{\vrule width 0.6pt height 4pt depth 0pt
              \hskip -3.5 pt} C}

\documentstyle[12pt]{article}
\begin{document}
\parskip=4pt
\parindent=18pt
\baselineskip=22pt \setcounter{page}{1} \centerline{\Large\bf On
Integrability of Many-Body Problems} \vspace{3ex}
\centerline{\Large\bf with Point Interactions} \vspace{5ex}

\centerline{S. Albeverio$^{1,2}$, S-M. Fei$^{1,3}$ and P.
Kurasov$^4$}

\parskip=0pt
\parindent=30pt
\baselineskip=16pt \vskip 1 true cm

$^1$ Institut f\"ur Angewandte Mathematik, Universit\"at Bonn,
D-53115 Bonn

$^2$ SFB 256; SFB 237; BiBoS; CERFIM (Locarno); Acc.Arch., USI
(Mendrisio)

$^3$ Max-Planck-Institute for Mathematics in the Sciences, 04103
Leipzig

~~Dept. of Math., Capital Normal University, Beijing 100037

$ ^4 $ Dept. of Math., Stockholm University, 10691 Stockholm,
Sweden

~~Dept. of Math., Lund Institute of Technology, 22100 Lund, Sweden

~~Dept. of Mathematical and Computational Physics, St. Petersburg

~~University, 198904 St. Petersburg, Russia

\parindent=18pt
\parskip=6pt

\begin{center}
\begin{minipage}{5in}
\vspace{3ex} \centerline{\large Abstract} \vspace{4ex} A study of
the integrability of one-dimensional quantum mechanical many-body
systems with general point interactions and boundary conditions
describing the interactions which can be independent or dependent
on the spin states of the particles is presented. The
corresponding Bethe ansatz solutions, bound states and scattering
matrices are explicitly given. Hamilton operators corresponding to
special spin dependent boundary conditions are discussed.
\end{minipage}
\end{center}
\vspace{5ex} \baselineskip=18pt

Exactly solvable models of a single quantum particle moving in a
local singular potential concentrated at one or a discrete number of points
have been extensively discussed in the literature, see e.g.
\cite{agh-kh,AKbook,gaudin} and references therein.
In the one dimensional case, the local singular potential
(contact interactions) at, say, the origin
($x=0$) can be characterized by
the boundary conditions imposed on the wave function $\varphi$ at
$x=0$. There are two classes of such boundary conditions:
separated and nonseparated boundary conditions, corresponding
to the cases where the perturbed operator is equal to
the orthogonal sum of two self--adjoint operators in $ L_{2}
(-\infty,0] $ and $L_{2} [0,\infty) $ and when this representation
is impossible, respectively.
The many-body problems with
pairwise interactions given by such boundary conditions
are generally not exactly solvable \cite{adf,afk,gauge}.
In the present paper we give a systematic description
for integrable models of many-body systems with
pairwise interactions given by such singular potentials
for the case where the boundary conditions
are independent as well as for the case where they are dependent
on the spin states of the particles.

We first consider the case of spin independent boundary conditions.
The family of point interactions for the one dimensional
Schr\"odinger operator $ - \frac{d^2}{dx^2}$
can be described by unitary $ 2 \times  2 $ matrices
via von Neumann formulas for self-adjoint extensions
of symmetric operators, since the second derivative
operator restricted to the domain $ C_{0}^\infty ({\bf R}
\setminus \{ 0 \} ) $ has deficiency indices $ (2,2)$.
The nonseparated boundary conditions
describing the self-adjoint extensions
have the following form
\begin{equation} \label{bound}
\left( \begin{array}{c}
\varphi\\
\varphi '\end{array} \right)_{0^+}
= e^{i\theta} \left(
\begin{array}{cc}
a & b \\
c & d \end{array} \right)
\left( \begin{array}{c}
\varphi\\
\varphi '\end{array} \right)_{0^-},
\end{equation}
where
\begin{equation}\label{abcd1}
ad-bc = 1,~~~~\theta, a,b,c,d \in \Rb.
\end{equation}
$\varphi(x)$ is the scalar wave function of two spinless particles with
relative coordinate $x$. (\ref{bound}) also describes two particles
with spin $s$ but without any spin coupling between the particles when they
meet (i.e. for $x=0$), in this case $\varphi$ represents any one of the
components of the wave function. The values $\theta = b=0$, $a=d=1$ in
(\ref{bound})
correspond to the case of a positive (resp. negative) $\delta$-function
potential for $c>0$ (resp. $c<0$). For general $a,b,c$ and $d$, the
properties of the corresponding Hamiltonian systems have been studied in
detail, see e.g. \cite{kurasov,ch,abd}.

The separated boundary conditions are described by
\begin{equation}\label{bounds}
\varphi^\prime(0_+) = q^+ \varphi (0_+)~, ~~~
\varphi^\prime(0_-) = q^- \varphi (0_-),
\end{equation}
where $q^{\pm} \in \Rb \cup \{ \infty\}$.
$q^+ = \infty$ or $q^- = \infty$
correspond to Dirichlet boundary conditions and
$q^+ = 0$ ~or~ $q^- = 0$ correspond to
Neumann boundary conditions.

To study the
integrability of one dimensional systems of $N$-identical particles
with general contact
interactions described by the boundary conditions
(\ref{bound}) or (\ref{bounds})
that are imposed on the relative coordinates of the particles,
we first consider the case of two particles ($N=2$) with
coordinates $x_1$, $x_2$ and momenta $k_1$, $k_2$ respectively.
Each particle has $n$-`spin' states designated by $s_1$ and $s_2$,
$1\leq s_i\leq n$. For $x_1\neq x_2$, these two particles are free. The
wave functions $\varphi$ are symmetric (resp. antisymmetric) with respect
to the interchange $(x_1,s_1)\leftrightarrow(x_2,s_2)$ for bosons (resp.
fermions). In the region $x_1<x_2$, from the Bethe ansatz the
wave function is of the form,
\begin{equation}\label{w1}
\varphi=u_{12}e^{i(k_1x_1+k_2x_2)}+u_{21}e^{i(k_2x_1+k_1x_2)},
\end{equation}
where $u_{12}$ and $u_{21}$ are $n^2\times 1$ column matrices.
In the region $x_1>x_2$, the wave function has the form
\begin{equation}\label{w2}
\varphi=(P^{12}u_{12})e^{i(k_1x_2+k_2x_1)}
+(P^{12}u_{21})e^{i(k_2x_2+k_1x_1)},
\end{equation}
where according to the symmetry or antisymmetry conditions,
$P^{12}=p^{12}$ for bosons and $P^{12}=-p^{12}$ for fermions, $p^{12}$
being the operator on the $n^2\times 1$ column that interchanges
$s_1\leftrightarrow s_2$.

Let $k_{12} = (k_1 -k_2)/2$. In the center of mass
coordinate $X=(x_1+x_2)/2$ and the relative coordinate
$x=x_2-x_1$, we get, by substituting (\ref{w1}) and (\ref{w2}) into the
boundary conditions at $x=0$,
\begin{equation}\label{2112}
u_{21} = Y_{21}^{12} u_{12}~,
\end{equation}
\begin{equation}\label{a21a12}
Y_{21}^{12}=\frac{2ie^{i\theta}k_{12}P^{12}+ik_{12}(a-d)+(k_{12})^2b+c}
{ik_{12}(a+d) + (k_{12})^2b-c}
\end{equation}
for boundary condition (\ref{bound}) and
\begin{equation}\label{a21a12s}
Y_{21}^{12}
=\frac{ik_{12} + q}{ik_{12} - q}
\end{equation}
for  boundary condition (\ref{bounds}), where
$q\equiv q_+ = - q_- \in \Rb \cup \{ \infty\}$.

For $N\geq 3$ and $x_1<x_2<...<x_N$, the wave function is given by
\begin{equation}\label{psi}
\begin{array}{rcl}
\psi&=&u_{12...N}e^{i(k_1x_1+k_2x_2+...+k_Nx_N)}
+u_{21...N}e^{i(k_2x_1+k_1x_2+...+k_Nx_N)}\\[3mm]
&&+(N!-2)~other~terms.
\end{array}
\end{equation}
The columns $u$ have $n^N\times 1$ dimensions. The wave functions
in the other regions are determined from (\ref{psi}) by the requirement of
symmetry (for bosons) or antisymmetry (for fermions).
Along any plane $x_i=x_{i+1}$, $i\in 1,2,...,N-1$, from similar
considerations as above we have
\begin{equation}\label{a1n}
\alpha_{\alpha_1\alpha_2...\alpha_i\alpha_{i+1}...\alpha_N}=Y_{\alpha_{i+1}\alpha_i}^{ii+1}
u_{\alpha_1\alpha_2...\alpha_{i+1}\alpha_i...\alpha_N},
\end{equation}
where
\begin{equation}\label{yn}
Y_{\alpha_{i+1}\alpha_i}^{ii+1}=
\frac{2ie^{i\theta}k_{\alpha_i\alpha_{i+1}}P^{ii+1}
+ik_{\alpha_i\alpha_{i+1}}(a-d) + (k_{\alpha_i\alpha_{i+1}})^2 b+c}
{ik_{\alpha_i\alpha_{i+1}}(a+d)+(k_{\alpha_i\alpha_{i+1}})^2 b-c}
\end{equation}
for nonseparated boundary condition and
\begin{equation}\label{ys}
Y_{\alpha_{i+1}\alpha_i}^{ii+1}=
\frac{ik_{\alpha_i\alpha_{i+1}} + q}{ik_{\alpha_i\alpha_{i+1}} - q}
\end{equation}
for separated boundary condition.
Here $k_{\alpha_i\alpha_{i+1}}=(k_{\alpha_i}-k_{\alpha_{i+1}})/2$ play the role of spectral
parameters.
$P^{ii+1}=p^{ii+1}$ for bosons and $P^{ii+1}=-p^{ii+1}$ for fermions,
with $p^{ii+1}$ the operator on the $n^N\times 1$ column
that interchanges $s_i\leftrightarrow s_{i+1}$.

For consistency $Y$ must satisfy the Yang-Baxter equation with
spectral parameter \cite{y,ma2,ma3,ma1,ma4}, i.e.,
$$
Y^{m,m+1}_{ij}Y^{m+1,m+2}_{kj}Y^{m,m+1}_{ki}
=Y^{m+1,m+2}_{ki}Y^{m,m+1}_{kj}Y^{m+1,m+2}_{ij},
$$
or
\begin{equation}\label{ybe11}
Y^{mr}_{ij}Y^{rs}_{kj}Y^{mr}_{ki}
=Y^{rs}_{ki}Y^{mr}_{kj}Y^{rs}_{ij}
\end{equation}
if $m,r,s$ are all unequal, resp.
\begin{equation}\label{ybe22}
Y^{mr}_{ij}Y^{mr}_{ji}=1,~~~~~~
Y^{mr}_{ij}Y^{sq}_{kl}=Y^{sq}_{kl}Y^{mr}_{ij}
\end{equation}
if $m,r,s,q$ are all unequal.

The operators $Y$ given by (\ref{yn}) satisfy the relation (\ref{ybe22})
for all $\theta ,a,b,c,d$. However the relations (\ref{ybe11}) are
satisfied only when $\theta =0$, $a=d$ and $b=0$, that
is, according to the constraint (\ref{abcd1}),
$\theta =0$, $a=d=\pm 1$, $b=0$, $c$ arbitrary.
The case $a=d=1$, $\theta =b=0$ corresponds to the usual $\delta$-function
interactions, which has been investigated in \cite{y,y2,y1}.
The case $a=d=-1$, $\theta =b=0$
is related to another singular interactions between any pair of particles
(for $a=d=-1$ and $\theta =b=c=0$ see \cite{kurasov,ch}),
which is in fact unitarily equivalent
to the $\delta$-interaction,
under a non-smooth ``kink type'' gauge transformation
$U = \prod_{i>j}{\rm ~sgn} (x_i-x_j)$.
Associated with the
separated boundary condition, the operators
$Y$ given by (\ref{ys}) satisfy both the
relations (\ref{ybe11}) and (\ref{ybe22}) for arbitrary $q$.
Therefore with respect to $N$-particle (either boson or
fermion) problems, there are two non-equivelant integrable one parameter
families with contact interactions
described respectively by one of the following conditions on the wave
function along the plane $x_i=x_j$ for any pair of particles
with coordinates $x_i$ and $x_j$,
\begin{equation}\label{b0}
\varphi(0_+)=+\varphi(0_-),~~~\varphi^\prime(0_+)=c\varphi(0_-)+\varphi^\prime(0_-)~, ~
c\in \Rb ~;
\end{equation}
\begin{equation}\label{bs}
\varphi^\prime (0_+)= q \varphi(0_+),~~~\varphi^\prime(0_-)= -q \varphi(0_-)~, ~
q\in \Rb \cup \{ \infty\}~.
\end{equation}
The wave functions are given by (\ref{psi}) with the
$u$'s determined by (\ref{a1n}) and initial conditions. The
operators $Y$ in (\ref{a1n}) are given respectively by
\begin{equation}\label{y0}
Y_{\alpha_{i+1}\alpha_i}^{ii+1}= \frac{i(k_{\alpha_i}-k_{\alpha_{i+1}})P^{ii+1} +c}
{i(k_{\alpha_i}-k_{\alpha_{i+1}}) - c} ~;
\end{equation}
and
\begin{equation}\label{y1s}
Y_{\alpha_{i+1}\alpha_i}^{ii+1}
=\frac{i(k_{\alpha_i}-k_{\alpha_{i+1}}) +2 q}{i(k_{\alpha_i}-k_{\alpha_{i+1}}) - 2 q} ~.
\end{equation}

When $q<0$, there exist $2^{N(N-1)/2}$ bound states for
the case (\ref{y1s}) of separated boundary conditions, with wavefunction
\begin{equation}\label{bpsins}
\psi_{N,\underline{\epsilon}}=
u_{\underline{\epsilon}}
\prod_{k>l} (\theta (x_k-x_l) +\epsilon_{kl}\theta (x_l-x_k))
e^{q\sum_{i>j} \vert x_i-x_j\vert }
\end{equation}
and eigenvalue $E=-q^2 N(N^2-1)/3$,
where $u_{\underline{\epsilon}}$ is the spin wave function and
$\underline{\epsilon} \equiv \{ \epsilon_{kl}~:~k>l \}$; $\epsilon_{kl}=\pm$,
labels the $2^{N(N-1)/2}$-fold degeneracy.

We consider now the case of spin dependent boundary conditions.
For a particle with spin $s$, the wave function has $n=2s+1$ components.
Therefore two particles with contact interactions have a general boundary
condition described in the center of mass coordinate system by:
\begin{equation}\label{BOUND}
\left( \begin{array}{c}
\psi\\
\psi '\end{array} \right)_{0^+}
=\left(
\begin{array}{cc}
A & B \\
C & D \end{array} \right)
\left( \begin{array}{c}
\psi\\
\psi '\end{array} \right)_{0^-},
\end{equation}
where $\psi$ and $\psi '$ are $n^2$-dimensional
column vectors, $A,B,C$ and $D$ are
$n^2\times n^2$ matrices. The boundary condition (\ref{BOUND})
can  include
not only the usual contact interaction between the particles,
but also a spin coupling
of the two particles if the matrices $ A, B, C, D $ are not diagonal.

The matrices $A,B,C$, and $D$ are subject to restrictions due to
the required symmetry condition of the Schr\"odinger operator.
For any $u,v\in C^\infty(\Rb \setminus \{0\})$,
$\displaystyle<-\frac{d^2}{dx^2}u,v>_{L_2(\Rb,\cb^n)}-
<u,-\frac{d^2}{dx^2}v>_{L_2(\Rb,\cb^n)}=0$, which, together with
(\ref{BOUND}) imply
\begin{equation}\label{ABCD}
A^\dagger D-C^\dagger B=1,~~~B^\dagger D=D^\dagger B,~~~
A^\dagger C=C^\dagger A,
\end{equation}
where $\dagger$ stands for the conjugate and transpose. Obviously
(\ref{bound}) is the special case of (\ref{BOUND}) for $s=0$.

In the following we study quantum systems with contact interactions
described by the boundary condition (\ref{BOUND}), in particular, $N$-body
systems with $\delta$-interactions.
We first consider two spin-$s$ particles with $\delta$-interactions.
The Hamiltonian is then of the form
\begin{equation}\label{H}
H=(-\frac{\partial^2}{\partial x_1^2}-\frac{\partial^2}{\partial
x_2^2}){\bf I}_2
+2h\delta(x_1-x_2),
\end{equation}
where ${\bf I}_2$ is the $n^2\times n^2$ identity matrix,
$h $
is an $n^2\times n^2$  Hermitian matrix.
If the matrix $h$ is proportional to the unit matrix
$ {\bf I}_{2}$, then $H$ is reduced to the usual
two-particle Hamiltonian with contact interactions but no spin
coupling.

Let $e_\alpha$, $\alpha=1,...,n$,
be the basis (column) vector with the $\alpha$-th component as $1$
and the rest components $0$. The wave function of the system (\ref{H})
is of the form
\begin{equation}\label{psis}
\psi=\sum_{\alpha,\beta=1}^n\phi_{\alpha
\beta}(x_1,x_2)e_\alpha\otimes e_\beta.
\end{equation}
In the center of mass coordinate system,
the operator (\ref{H}) has the form
\begin{equation}
H = - \left( \frac{1}{2} \frac{\partial^2}{\partial X^2}
+ 2 \frac{\partial ^2}{\partial x^2} \right)
{\bf I}_{2} + 2 h \delta (x).
\end{equation}
The functions $ \phi = \phi (x,X) $
 from the domain of this operator
satisfy
the following boundary condition at $x=0$,
\begin{equation}\label{b2}
\phi_{\alpha\beta} '(0^+,X)-\phi_{\alpha\beta} '(0^-,X)=
\sum_{\alpha,\beta=1}^n h_{\gamma\lambda,\alpha\beta}
\phi_{\gamma\lambda}(0,X),~~~
\phi_{\alpha\beta}(0^+,X)=\phi_{\alpha\beta}(0^-,X),
\end{equation}
$\alpha,\beta=1,...,n$,
where the indices of the matrix $h$ are arranged as $11,12,...,1n$; $21,22,...,
2n$; ...; $n1,n2,...,nn$. (\ref{b2}) is a special case of (\ref{BOUND}) for
$A=D={\bf I}_2$, $B=0$ and $C=h$. $h$ acts on the basis vector of particles $1$
and $2$ by $h e_\alpha\otimes e_\beta =\displaystyle
\sum_{\gamma,\lambda=1}^n h_{\alpha\beta,\gamma\lambda} e_\gamma\otimes
e_\lambda$.

The wave functions are still of the forms (\ref{w1})
(resp. \ref{w2}) in the region $x_1<x_2$ (resp. $x_1>x_2$).
Substituting them into the boundary
conditions
(\ref{b2}), we get
\begin{equation}\label{a1}
\left\{
\begin{array}{l}
u_{12}+u_{21}
=P^{12}(u_{12}+u_{21}),\\
ik_{12}(u_{21}-u_{12})
=hP^{12}(u_{12}+u_{21})+ik_{12}P^{12}(u_{12}-u_{21}).
\end{array}\right.
\end{equation}
Eliminating the term $P^{12}u_{12}$ from
(\ref{a1}) we obtain the same relation as (\ref{2112}),
$u_{21} = Y_{21}^{12} u_{12}$. Nevertheless the $Y$ operator is given by
\begin{equation}\label{a21a12ss}
Y_{21}^{12}=[2ik_{12} -h]^{-1}[2ik_{12}P^{12}+h].
\end{equation}

For a system of $N$ identical particles with $\delta$-interactions,
the Hamiltonian is given by
\begin{equation}\label{HN}
H=-\sum_{i=1}^N\frac{\partial^2}{\partial x_i^2}{\bf I}_N+\sum_{i<j}^N h_{ij}
\delta(x_i-x_j),
\end{equation}
where ${\bf I}_N$ is the $n^N\times n^N$ identity matrix,
$h_{ij}$ is an operator acting on the $i$-th and $j$-th bases as $h$ and the
rest as identity, e.g., $h_{12}=h\otimes {\bf 1}_3\otimes...{\bf 1}_N$,
with ${\bf 1}_i$ the $n\times n$ identity matrix acting on the $i$-th basis.
The wave function in a given region, say $x_1<x_2<...<x_N$, is of the form
(\ref{psi}), with
\begin{equation}\label{a1ns}
u_{\alpha_1\alpha_2...\alpha_j\alpha_{j+1}...\alpha_N}=
Y_{\alpha_{j+1}\alpha_j}^{jj+1}
u_{\alpha_1\alpha_2...\alpha_{j+1}\alpha_j...\alpha_N}
\end{equation}
and
\begin{equation}\label{y}
Y_{\alpha_{j+1}\alpha_j}^{jj+1}=[2ik_{\alpha_j\alpha_{j+1}}-h_{j{j+1}}]^{-1}
[2ik_{\alpha_j\alpha_{j+1}}P^{jj+1} + h_{jj+1}].
\end{equation}

From the Yang-Baxter equations it is
straightforward to show that the operator $Y$ given
by (\ref{y}) satisfies all the Yang-Baxter relations if
\begin{equation}\label{hp}
[h_{ij}, P^{ij}]=0.
\end{equation}
Therefore if the Hamiltonian operators for the spin coupling commute with the
spin permutation operator, the $N$-body quantum system (\ref{HN}) can be
exactly solved. The wave function is then given by (\ref{psi}) and
(\ref{a1ns}) with the energy $E=\displaystyle\sum_{i=1}^N k_i^2$.

For the case of spin-$1\over 2$, a Hermitian
matrix satisfying (\ref{hp}) is generally of the form
\begin{equation}\label{hspin}
h^{1\over 2}=\left(\ba{cccc}a&e_1&e_1&c\\
e_1^\ast&f&g&e_2\\e_1^\ast&g&f&e_2\\
c^\ast&e_2^\ast&e_2^\ast&b\ea\right),
\end{equation}
where $a,b,c,f,e_1,e_2\in\Cb$, $g\in \Rb$.
We recall that for a complex vector space  $V$, a matrix $R$
taking values in $End_c(V\otimes V)$ is called a solution of the
Yang-Baxter equation without spectral parameters, if it satisfies
$R_{12}R_{13}R_{23}=R_{23}R_{13}R_{12}$,
where $R_{ij}$ denotes the matrix on the complex vector space
$V\otimes V\otimes V$, acting as $R$ on the $i$-th and the $j$-th
components and as identity on the other components. When $V$ is a two
dimensional complex space, the solutions of the Yang-Baxter equation include
the ones such as $R_q$ which gives rise to the quantum algebra
$SU_q(2)$ and the integrable Heisenberg spin-$1\over 2$ chain models
such as the XXZ model ($R$ corresponds
to the spin coupling operator between the nearest neighbor spins in
Heisenberg spin chain models)\cite{ma2,ma3,ma1,ma4}.
Nevertheless in general $h^{1\over 2}$ does not satisfy the
Yang-Baxter equation without spectral parameters: $h^{1\over
2}_{12}h^{1\over 2}_{13}h^{1\over 2}_{23}\neq h^{1\over 2}_{23}h^{1\over
2}_{13}h^{1\over 2}_{12}$. But (\ref{hspin}) includes the Yang-Baxter
solutions, such as $R_q$, that
gives integrable spin chain models
(for an extensive investigation of the Yang-Baxter solutions see
\cite{ybe,jamo}). Therefore for an $N$-body system to be integrable,
the spin coupling in the contact interaction (\ref{HN}) is allowed to be
more general than the spin coupling in a Heisenberg spin chain model with
nearest neighbors interactions.

For $N=2$, from (\ref{a1}) the bound states have the form,
\begin{equation}\label{bpsi2}
\psi^2_\alpha=u_\alpha e^{\frac{c+a\Lambda_\alpha}{2}\vert x_2-
x_1\vert},~~~~\alpha=1,...,n^2,
\end{equation}
where $u_\alpha$ is the common $\alpha$-th eigenvector of $h$ and
$P^{12}$, with eigenvalue $\Lambda_\alpha$,
s.t. $hu_\alpha=\Lambda_\alpha u_\alpha$ and $c+a\Lambda_\alpha<0$,
$P^{12}u_\alpha=u_\alpha$.
The eigenvalue of the Hamitonian $H$ corresponding to the bound state
(\ref{bpsi2})
is $-(c+a\Lambda_\alpha)^2/2$. We remark that, whereas for the case of
the boundary condition (\ref{bound}), for a $\delta$ interaction one has
a unique bound state, here we have $n^2$ bound states.
By generalization we get the bound state for the $N$-particle system,
\begin{equation}\label{bpsin}
\psi^N_\alpha=v_\alpha e^{-\frac{c+a\Lambda_\alpha}{2}\sum_{i>j}\vert
x_i-x_j\vert},~~~\alpha=1,...,n^2,
\end{equation}
where $v_\alpha$ is the wave function of the spin part
satisfying $P^{ij}v_\alpha=v_\alpha$ and
$h_{ij}v_\alpha=\Lambda_\alpha v_\alpha$, for any $i\neq j$.

It is worth mentioning that $\psi^N_\alpha$ is of the form (\ref{psi})
in each of the above regions. For instance comparing $\psi^N_\alpha$
with (\ref{psi}) in the region $x_1<x_2...<x_N$ we get
\begin{equation}\label{k}
k_1=-i\frac{c+a\Lambda_\alpha}{2}(N-1),~k_2=k_1+ic,~k_3=k_2+ic,...,k_N=-k_1,
\end{equation}
for $\alpha=1,...,n^2$. The energy of the bound state $\psi^N_\alpha$ is
\begin{equation}\label{e}
E_\alpha=-\frac{(c+a\Lambda_\alpha)^2}{12}N(N^2-1).
\end{equation}

Now we pass to the scattering matrix. For real $k_1<k_2<...k_N$, in
each coordinate region such as $x_1<x_2<...x_N$, the following term in
(\ref{psi}) describes an outgoing wave
$\psi_{out}=u_{12...N}e^{i(k_1x_1+...+k_Nx_N)}$.
An incoming wave with the same exponential as $\psi_{out}$ is given by
$\psi_{in}=[P^{1N}P^{2(N-1)}...]u_{N(N-1)...1}e^{i(k_Nx_N+...+k_1x_1)}$
in the region $x_N<x_{N-1}<...<x_1$. From (\ref{a1ns})
the scattering matrix $S$ defined by $\psi_{out}=S\psi_{in}$ is given by
$S=[X_{21}X_{31}...X_{N1}][X_{32}X_{42}...X_{N2}]...[X_{N(N-1)}]$,
where $X_{ij}=Y^{ij}_{ij}P^{ij}$.

The scattering matrix $S$ is unitary and
symmetric due to the time reversal invariance of the interactions.
$<s_1^\p s_2^\p...s_N^\p\vert S\vert s_1s_2...s_N>$ stands for the $S$
matrix element of the process from the state
$(k_1s_1,k_2s_2,...,k_Ns_N)$ to the
state $(k_1s_1^\p,k_2s_2^\p,...,k_Ns_N^\p)$.

The scattering of clusters (bound states)
can be discussed in a similar way as in \cite{y1}.
For instance for the scattering of a bound state of two
particles ($x_1<x_2$) on a bound state of three particles
($x_3<x_4<x_5$), the scattering matrix is $S=[X_{32}X_{42}X_{52}]
[X_{31}X_{41}X_{51}]$.

The integrability of many particles systems with contact spin coupling
interactions governed by separated boundary conditions can also be
studied. Instead of (\ref{BOUND}) we need to deal with the case
\begin{equation}\label{BOUNDS}
\phi^\prime(0_+) = G^+ \phi (0_+), ~~~
\phi^\prime(0_-) = G^- \phi (0_-) ,
\end{equation}
where $G^{\pm}$ are Hermitian matrices.
For $G^+=G^-\equiv G$, $G^\dagger =G$, there is a Bethe Ansatz solution
to (\ref{psi}) with $Y_{\alpha_{i+1}\alpha_i}^{ii+1}$ in (\ref{a1ns}) given by
\begin{equation}\label{yss}
Y_{\alpha_{i+1}\alpha_i}^{ii+1}=
\frac{ik_{\alpha_i\alpha_{i+1}} + G}{ik_{\alpha_i\alpha_{i+1}} - G}.
\end{equation}

Let $\Gamma$ be the set of $n^2$ eigenvalues of $G$. For any
$\lambda_\alpha\in \Gamma$ such that $\lambda_\alpha<0$, there are
$2^{N(N-1)/2}$ bound states for the $N$-particle system,
\begin{equation}\label{bpsinss}
\psi^{N}_{\alpha\underline{\epsilon}}=
v_{\alpha\underline{\epsilon}}
\prod_{k>l} (\theta (x_k-x_l) +\epsilon_{kl}\theta (x_l-x_k))
e^{\lambda_\alpha\sum_{i>j} \vert x_i-x_j\vert },
\end{equation}
where $v_{\alpha\underline{\epsilon}}$ is the spin wave function and
$\underline{\epsilon} \equiv \{ \epsilon_{kl}~:~k>l \}$; $\epsilon_{kl}=\pm$,
labels the $2^{N(N-1)/2}$-fold degeneracy.
The spin wave
function $v$ here satisfies $P^{ij}v_{\alpha\underline{\epsilon}}
=\epsilon_{ij}v_{\alpha\underline{\epsilon}}$
for any $i\neq j$, that is, $p^{ij}v_{\alpha\underline{\epsilon}}
=\epsilon_{ij}v_{\alpha\underline{\epsilon}}$
for bosons and $p^{ij}v_{\alpha\underline{\epsilon}}
=-\epsilon_{ij}v_{\alpha\underline{\epsilon}}$ for fermions.

Again $\psi^{N}_{\alpha\underline{\epsilon}}$ is of the form (\ref{psi})
in each of the regions $x_{i_1}<x_{i_2}<...<x_{i_N}$.
For instance comparing $\psi^{N}_{\alpha\underline{\epsilon}}$ with
(\ref{psi}) in the region $x_1<x_2...<x_N$ we get
$k_1=i\lambda_\alpha (N-1)$, $k_2=k_1-2i\lambda_\alpha$,
$k_3=k_2-2i\lambda_\alpha$,...,$k_N=-k_1$.
The energy of the bound state $\psi^{N}_{\alpha\underline{\epsilon}}$ is
$E_\alpha=-\lambda_\alpha^2N(N^2-1)/3$.

We have investigated the integrable models of $N$-body systems with
contact spin coupling interactions. Without taking into account the spin
coupling, the boundary condition (\ref{bound}) is characterized by four
parameters (separated boundary conditions are a special limiting
case of these). Obviously the general boundary condition (\ref{BOUND}) we
considered in this article has much more parameters. A complete
classification of the dynamic operators associated with
different parameter regions remains to be done. As we
have seen, the case $A=D={\bf I}_2$, $B=0$, $C=h$
corresponds to a Hamiltonian with $\delta$-interactions of the form
(\ref{H}) (for $N=2$). It can
be further shown that (for $N=2$) the following boundary condition
\begin{equation}\label{BOUND1}
\left( \begin{array}{c}
\psi\\
\psi '\end{array} \right)_{0^+}
=\left(
\begin{array}{cc}
{\bf I} & B \\
0 & {\bf I} \end{array} \right)
\left( \begin{array}{c}
\psi\\
\psi '\end{array} \right)_{0^-}
\end{equation}
corresponds to a Hamiltonian $H$ of the form:
$H=-D^2_x(1+B\delta) - BD_x\delta^\prime$,
where $B$ is an $n^2\times n^2$ Hermitian matrix,
$D_x$ is defined by $(D_x f)(\varphi)=-f(\frac{d}{dx}\varphi)$, for
$f\in C^\infty_0(\Rb\slash\{0\})$ and $\varphi$ a test function with a
possible discontinuity at the origin. The boundary condition
\begin{equation}\label{BOUND2}
\left( \begin{array}{c}
\psi\\
\psi '\end{array} \right)_{0^+}
=\left(
\begin{array}{cc}
\frac{2+iB}{2-iB} & 0 \\
0 & \frac{2-iB}{2+iB} \end{array} \right)
\left( \begin{array}{c}
\psi\\
\psi '\end{array} \right)_{0^-}
\end{equation}
describes the Hamiltonian
$H=-D_x^2+iB(2D_x\delta-\delta^\prime)$.

\vspace{2.5ex}
\noindent
ACKNOWLEDGEMENTS: S.M. Fei would like to thank DFG, Max-Planck-Institute
for Mathematics in the Sciences, Leipzig for financial supports and
warm hospitalities.

\end{document}